# Self-assembly of liquid crystal block copolymer PEG-*b*-smectic polymer in pure state and in dilute aqueous solution

Bing Xu,[a,†] Rafael Piñol,[a,‡] Merveille Nono-Djamen,[a,b#] Sandrine Pensec,[b] Patrick Keller,[a] Pierre-Antoine Albouy,[c] Daniel Lévy[a] and Min-Hui Li*[a]

[a] Institut Curie, Centre de Recherche; CNRS, UMR 168; Université Pierre et Marie Curie; 75248 Paris CEDEX 05, France.

[b] Université Pierre et Marie Curie, CNRS UMR7610, Laboratoire de Chimie des Polymères, Case 185, 4, Place Jussieu, 75252 Paris CEDEX 05, France.

[c] Université Paris-Sud, CNRS UMR8502, Laboratoire de Physique des Solides, 91405 Orsay CEDEX, France.

* Corresponding author: Institut Curie, 26 rue d'Ulm, Paris, France.
Fax: 33 1 4051 0636; Tel: 33 1 5624 676; E-mail: min-hui.li@curie.fr

**Summary**

A series of amphiphilic LC block copolymers, in which the hydrophobic block is a smectic polymer poly(4-methoxyphenyl 4-(6-acryloyloxy-hexyloxy)-benzoate) (PA6ester1) and the hydrophilic block is polyethyleneglycol (PEG), were synthesized and characterized. The self-assembly of one of them in both the pure state and the dilute aqueous solution was investigated in detail. Nano-structures in the pure state were studied by SAXS and WAXS on samples aligned by a magnetic field. A hexagonal cylindrical micro-segregation phase was observed with a lattice distance of 11.2 nm. The PEG blocks are in the cylinder, while the smectic polymer blocks form a matrix with layer spacing 2.4 nm and layer normal parallel to the long axis of the cylinders. Faceted unilamellar polymer vesicles, polymersomes, were formed in water, as revealed by cryo-TEM. In the lyotropic bilayer membrane of these polymersomes, the thermotropic smectic order in the hydrophobic block is clearly visible with layer normal parallel to the membrane surface.

# 1. Introduction

Nanostructures formed by block copolymers composed of immiscible blocks have been extensively studied theoretically and experimentally.[1,2] It is now well established that these block copolymers can phase separate into a variety of organized structures (lamellae, cylinder, double gyroid, spherical etc.) in the solid state,[3] and form a variety of micellar aggregates (spherical and cylindrical micelles, vesicles etc.) when dissolved in a solvent selective for one of the blocks.[4] These materials could find potential applications in various fields of nanotechnology, such as nano soft-lithography, nanoreactors and drug delivery systems.[5] These applications require, however, that these materials possess additional properties beyond those inherent to their intrinsic nanostructures. Perfect periodic nanodomain ordering is necessary, for example, for nano soft-lithography, while response to stimuli is desirable for nanoreactors and drug delivery systems.

In the past decade, electric fields,[6] temperature gradients,[7] directional solidification,[8] modification of substrate surfaces,[9] shearing,[10] solvent evaporation,[11] and roll casting[12] have been explored to induce orientation and/or long-range order in the nanostructures of classical coil-coil block copolymers in the solid state. A perfect long-range pattern of such nanostructures by non-contacted and remote approaches still remains however a challenge. Recent progress in synthetic strategies simplifies the molecular design of block copolymers. Block copolymers with a variety of chemical compositions and architectures can be prepared by synthetic tools developed in macromolecular chemistry, including living and controlled polymerization techniques. One of the ideas then is to develop block copolymers which possess intrinsic responses to external remote triggers. These responses should help achieve nanostructure control. Liquid crystalline (LC) rod-coil block copolymers are good candidates because they respond to multiple stimuli including electric and magnetic fields, temperature and light (if the mesogen is a chromophore). LC rod-coil block copolymers exhibit hierarchical order in the solid state: microphase segregation in the block copolymer may occur on the 10-100 nm length scale, while LC ordering is observed on the length scale of mesogen (typically 2-4 nm). These two levels of order, in the block copolymer and liquid crystalline mesophases, interplay with each other.[13,14,15] Non-contacted and remote approaches, such as magnetic fields[16,17,18] and light,[19,20] have recently been used to align the nanostructures of LC block copolymers.

When the coil block is hydrophilic, an amphiphilic LC rod–coil block copolymer results representing a new class of amphiphilic copolymers. These copolymers form interesting micellar aggregates in aqueous solution, including vesicles and nanofibers.[21,22] Much effort has been made recently to develop stimuli-responsive polymer micelles and vesicles[23,24] because of their potential applications in nanoreactors and drug delivery systems. Key chanllenge is the controlled release of the active substances in the micellar aggregates, both in space and time. Most of the developed systems make use of chemical stimuli, which require the addition of chemical reagents, and are not always compatible with the application environments. In contrast, LC block copolymers are a promising system for the development of responsive micelles and vesicles sensitive to non-contacted and remote physical stimuli. We have shown recently that the use of light-sensitive LC nematic block copolymers allows the creation of polymer vesicles which burst under UV illumination.[25]

As a part of our endeavour to develop well-defined and stimuli-responsive nanostructures made from LC block copolymers, we report in this paper the synthesis and characterization of a series of amphiphilic LC block copolymers, in which the hydrophobic block is a smectic polymer and the hydrophilic block is polyethyleneglycol (PEG). The self-assembly of one of the LC block copolymers in the pure state and in dilute aqueous solution was then investigated in detail. A well aligned cylindrical nanostructure was obtained in the solid state using a magnetic field to induce orientation in the LC block. In water, faceted unilamellar polymer vesicles (polymersomes) with smectic order in the lyotropic bilayer were formed.

**2. Experimental**

**2.1. Materials and analytical methods**

For monomer synthesis, 4-hydroxybenzoic acid, 6-chlorohexanol, KOH, KI, acrylic acid, *p*-toluenesulfonic acid, hydroquinone, 4-methoxyphenol, 4-pyrrolidinopyridine and dicyclohexylcarboxydiimide of analytical grade were purchased from Aldrich and used without further purification. Anhydrous $CH_2Cl_2$ and $CHCl_3$, as well as ethanol and isopropanol, were purchased from Carlo Erba-SDS. For the polymer synthesis, methoxy poly(ethylene)glycol (MPEG2000, $M_n$=2000, DP=45, from Fluka) was dried by azeotropic distillation with toluene before use, and traces of residual toluene were removed under vacuum. CuBr was purified by stirring with acetic acid for several hours, then filtrating, washing with acetic acid, ethanol and diethyl ether in succession and stored under vacuum. 4,4'-di(*n*-nonyl)-2,2'-bipyridine (from Aldrich) was recrystallized twice from ethanol. Other reagents (ethyl 2-bromo-2-methylpropionate, 2-bromopropionyl bromide and triethylamine from Aldrich) and solvents of analytical grade were used as received without further purification.

Molecular structures of all products were analyzed by $^1$H-NMR using a Bruker HW300MHz spectrometer. Molecular weight distributions ($M_w/M_n$) of polymers were evaluated by size exclusion chromatography (SEC) calibrated with polystyrene standards.[26] For SEC, we used Waters Styragel HR5E columns and a Waters 410 differential refractometer with THF as eluent at a flow rate of 1.0 mL/min at 40°C. Molecular weights of the LC homopolymer and the diblock copolymers were calculated from the NMR signals, using the following equations. For the LC homopolymer,

$$M_n = 195 + n \times 398,$$

$$n = \frac{3}{8} \times \frac{I_{aryl}}{I_{1.22}},$$

where $I_{aryl}$ denotes the integration of peaks of aryl hydrogens from 6.91 to 8.14 ppm and $I_{1.22}$ that of the methyl hydrogens in the chain end (see Scheme 1). n is the degree of polymerization (DP) of the LC polymer. The molecular weight of the monomer is 398 and that of initiator 195. For block copolymers,

$$M_n = M_n (MPEG-Br) + n \times 398$$

$$n = \frac{3}{8} \times \frac{I_{aryl}}{I_{3.38}}$$

where $I_{3.38}$ denotes the integration of peaks of terminal methyl hydrogens in MPEG.

**2.2. Synthesis of the LC monomer and homopolymer**

The LC acrylate monomer, 4'-methoxyphenyl 4-(6"-(acryloyloxy)hexyloxy)benzoate (A6ester1) was synthesized from 4-hydroxybenzoic acid by a three-step procedure as described in reference.[27] The monomer A6ester1, purified by recrystallization from ethanol (5 times), was obtained as white crystals ready for polymerization. $^1$H NMR (300

MHz, CDCl$_3$): δ 1.47-1.86 (m, 8H, -OCH$_2$(C$\underline{H}_2$)$_4$CH$_2$O-), δ 3.82 (s, 3H, -OC$\underline{H}_3$), δ 4.02-4.06 (t, 2H, -COOC$\underline{H}_2$CH$_2$-), δ 4.16-4.20 (t, 2H, -CH$_2$C$\underline{H}_2$OC$_6$H$_4$-), δ 5.80-5.83 (d, 1H, -CH=$\underline{H}$CH), δ 6.07-6.17 (m, 1H, -C$\underline{H}$=CH$_2$), δ 6.37-6.43 (d, 1H, -CH=HC$\underline{H}$), δ 6.91-8.14 (m, 8 H arom.)

The homopolymer was synthesized by atom transfer radical polymerization (ATRP). A general ATRP procedure is as follows:

A Schlenk flask with a magnetic stir bar was charged with Cu$^I$Br (14.34 mg, 0.1 mmol), 4,4'-di(*n*-nonyl)-2,2'-bipyridine (bpy9, ligand) (81.76 mg, 0.2 mmol), ethyl 2-bromo-2-methylpropionate (I, initiator) (19.51 mg, 0.1 mmol) and monomer A6ester1 (M) (0.4776 g, 1.2 mmol). The flask was degassed by four vacuum-argon cycles. Toluene (1 mL), degassed by argon bubbling for 30 min, was then introduced into the flask using a syringe purged with argon. The flask was then immersed in an oil bath held at 80 °C by means of a thermostat. After reacting for 24 h, the mixture was cooled to room temperature. The resulting polymer solution was poured into a large volume of diethyl ether. The precipitated polymer was purified thrice by dissolution in a small amount of dichloromethane and precipitation into a large volume of diethyl ether. The purified polymer was dried under vacuum at room temperature for 3 days. Yield: 0.2g (42%). $^1$H NMR (300 MHz, CDCl$_3$): δ 1.11-1.14 (6H, -C(C$\underline{H}_3$)$_2$), δ 1.21 (t, 3H, -CH$_2$C$\underline{H}_3$), δ 1.32-1.88 (8*n*H, -OCH$_2$(C$\underline{H}_2$)$_4$CH$_2$O-), δ 2.33 (*n*H, -CH$_2$C$\underline{H}$(COO)-), δ3.80 (3*n*H -OC$\underline{H}_3$), δ 3.91-4.18 (4*n*H+2H, -COOC$\underline{H}_2$CH$_2$-, -CH$_2$C$\underline{H}_2$OC$_6$H$_4$- and -

COOC$\underline{H}_2$CH$_3$), δ 6.82-8.15 (8$n$H arom.) ($n$ is DP of LC polymer calculated by $^1$H NMR spectrum). M$_n$ of PA6ester1 was 3300 (NMR) and M$_w$/M$_n$ = 1.17 (SEC).

**2.3. Synthesis of block copolymers PEG-*b*-smectic polymer**

The synthesis of the macroinitiator MPEG2000-Br (I) was first carried out as previously described.[21] Yield 58%. $^1$H NMR (300 MHz, CDCl$_3$): δ 1.81(d, J=6Hz, 3H, BrCH($\underline{CH}_3$)-), 3.36(s, 3H, $\underline{CH}_3$OCH$_2$CH$_2$O-), 3.51~3.87(m, 181H, -O$\underline{CH}_2\underline{CH}_2$O-), 4.30(t, J=3Hz, 2H, -CH$_2$$\underline{CH}_2$OCO-), 4.38(q, J=9Hz, 1H, Br$\underline{CH}$(CH$_3$)-).

The block copolymers PEG-*b*-PA6ester1 were synthesized by ATRP by the same procedure and conditions used for homopolymer PA6ester1, except that the macroinitiator MPEG2000-Br was used here instead of ethyl 2-bromo-2-methylpropanoate. The molar amount of the initiator was kept to be [I] = 0.1 mmol, and the molar ratio [I]/[bpy9]/[CuBr] = 1/2/1. The monomer molar ratio was varied as [M]/[I] = 5/1, 10/1, 15/1, 20/1 and 30/1, in order to obtain different lengths for the LC block (see Table 1). For a typical synthesis of copolymer CP3 (Table 1), 597.7 mg monomer A6ester1 (1.5 mmol) and 213.5 mg MPEG2000-Br (0.1 mmol) were polymerized according to the procedure described above, yielding 578.7 mg diblock copolymer (71.4%). $^1$H NMR (300 MHz, CDCl$_3$): δ 1.29-2.02 (8$n$ H, -CH$_2$($\underline{CH_2}$)$_4$CH$_2$O-), δ 2.31 ( $n$ H, -CH$_2$$\underline{CH}$(COO)-), δ 3.37 (3H, -O$\underline{CH}_3$), δ 3.57-3.57 (181 H –O$\underline{CH_2CH_2}$O-), δ 3.77 (3*n* H -O$\underline{CH}_3$), δ 3.89-4.15 (4*n*H+2H, -COO$\underline{CH}_2$CH$_2$-, -CH$_2$$\underline{CH}_2$OC$_6$H$_4$- and -COO$\underline{CH}_2$CH$_3$), δ 6.79-8.13 (8*n* H, arom.) (*n* is the DP of the LC block determined by the $^1$H NMR spectrum) $M_n$(CP3) = 7900 (NMR) and $M_w/M_n$ = 1.13 (SEC).

**2.5. Physical characterization**

The mesomorphic properties of the diblock copolymers in bulk were studied by thermal polarizing optical microscopy (POM) using a Leitz Ortholux microscope equipped with a Mettler FP82 hot stage, and differential scanning calorimetry using a Perkin-Elmer DSC7.

The self-assembled phases of the diblock copolymers in solid state were studied by X-ray scattering using CuK$_\alpha$ radiation ($\lambda$ = 1.54 Å) from a 1.5 kW rotating anode generator. The diffraction patterns were recorded on photosensitive imaging plates. The experiments were performed on fibre samples drawn from molten polymer for LC homopolymer and on samples contained in glass capillaries (1 mm diameter) for block copolymer. WAXS and SAXS were performed in two separate apparati. WAXS experiments examined the wave vector domain ($q$ = 4πsinθ/λ) from 1.83 – 32 nm$^{-1}$ and SAXS experiments probed the range 0.36 – 2.96 nm$^{-1}$. For the diblock copolymers, samples were first aligned by slowly cooling (0.1°C min$^{-1}$) from the isotropic phase (T = 85°C) to smectic phase (T = 40°C) in a magnetic field of 1.7 T. The magnetic field (**H**) was perpendicular to the long axis of the capillary containing the block copolymers. WAXS experiments on aligned samples were then made *in-situ* in the smectic phase (40°C), followed by SAXS experiments on the same sample cooled to room temperature. The X-ray was transverse to the plane formed by the magnetic field (**H**) and the capillary axis.

The preparation of polymer vesicles and the turbidity measurements were performed according to published procedures.[21] The diblock copolymers were first dissolved in dioxane, which is a good solvent for both polymer blocks, at a concentration of 1.0 wt%. Deionized water was then added very slowly to the solution (2-3 µL of water per minute to 1 mL of polymer solution) under slight shaking. After each addition of water, the solution was left to equilibrate for 10 or more minutes until the optical density was stable. The optical density (turbidity) was measured at a wavelength of 650 nm using a quartz cell (path length: 2 cm) with a Unicam UV/Vis spectrophotometer. The cycle of water addition, equilibration and turbidity measurement was continued until the turbidity reaches a plateau. The solution was then dialyzed against water for 3 days to remove dioxane using a Spectra/Por regenerated cellulose membrane with a molecular weight cut-off of 3500. The morphological analysis of the turbid polymer solutions was performed by cryo-TEM on samples flash frozen in liquid ethane. Images were recorded in low dose conditions using a Philips CM120 electron microscope equipped with a Gatan SSC 1Kx1K CCD camera. The calibration of the microscope was performed with purple membrane leading to 0.386 nm/pixel at 45 K magnification.

## 3. Results and discussion

### 3.1. Synthesis

LC homopolymer and block copolymers were prepared by ATRP as detailed in the experimental section (see Scheme 1 and Scheme 2). The same hydrophilic macroinitiator (MPEG2000-Br) was used to synthesize a series of block copolymers. By varying the monomer to macroinitiator molar ratio [M]/[I] from 5/1 to 30/1, one obtains LC hydrophobic blocks with different DP. Block copolymers with hydrophilic/hydrophobic weight ratios from 50/50 to 14/86 were synthesized. The SEC measurement showed narrow molecular distributions for all polymer samples ($M_w/M_n$ = 1.07 – 1.32). In Table 1 we summarize the molecular characteristics of the LC homopolymer and block copolymers analysed by $^1$H NMR and SEC. In this paper we will focus on the self-assembling properties of one of the block copolymers, $PEG_{45}$-*b*-$PA6ester1_{20}$ (CP4, hydrophilic/hydrophobic weight ratio of 20/80) in both solid state and dilute aqueous solution.

### 3.2. Mesomorphic properties

Before discussion of the self-assembly of block copolymers, we examine the individual mesomorphic properties of the LC monomer, homopolymer and copolymer CP4 respectively. The LC monomer A6ester1 has a monotropic nematic (N) phase with the phase sequence: Cr-61.9°C-I on heating, and I-47.7°C-N 35.4°C-Cr on cooling, as observed optically with a temperature change rate of 5°C/min (Cr represents the crystalline phase and I the isotropic phase). Typical Schlieren-type textures were observed in the nematic phase by POM. Once polymerized, the nematic A6ester1 gave a homopolymer (PA6ester1) with a richer polymorphism. PA6ester1 has a nematic phase at high temperature and a smectic phase at low temperature, in agreement with results published previously for the homopolymer with the same monomer structure, but much higher molecular weight.[27] Fig. 1a shows nematic Schlieren-type textures taken at T = 84.8°C and Fig. 1b shows fan-shape textures, typical for a smectic A (SmA) phase, taken at T = 75.0°C. The mesomorphic properties of PA6ester1, as determined by DSC, are shown in Fig. 2 and Table 2. The SmA phase is further confirmed by SAXS. Fig. 3 shows the SAXS pattern obtained on an aligned fiber sample of PA6ester1. The signals along the meridian give a period p = 2.52 nm for this smectic A phase. The extended molecular length of the side group is estimated to be 2.35 – 2.45 nm. A one-layer anti-parallel packing is therefore the most probable arrangement for the SmA mesophase.[28]

For the block copolymer sample CP4 the nematic and smectic mesophases seem to be preserved in the LC block, as suggested by the DSC curves where the same number of peaks were detected for CP4 and PA6ester1 (Fig. 2). The transition domains broaden and transition temperatures drop because of the presence of the amorphous PEG block.[21,26] The POM textures are Schlieren-type between 74 and 60 °C when cooling from the isotropic phase, indicating a nematic phase in this temperature range. At lower temperatures the fan-type textures are observed only after long annealing time. The final smectic phase assignment at T < 60°C is made by WAXS on samples aligned by a magnetic field.

The driving force for the magnetic field alignment of mesophases composed of typical aromatic LC mesogens is the collective anisotropy of the mesophase diamagnetic susceptibility, $\chi_\alpha = \chi_{//} - \chi_\perp$, with respect to the direction of the long axis of the LC mesogen (director $n$).[29] This anisotropy causes a free energy difference between the state of randomly oriented polydomains and the state of parallel- or perpendicular-oriented monodomains. For typical aromatic LC mesogens this difference is sufficiently large compared to typical thermal energy, $kT$, that the system forms a monodomain mesophase as prescribed by the diamagnetic anisotropy. The mesophase with positive anisotropy ($\chi_\alpha > 0$) will have the director $n$ aligned parallel to the field direction H, while the mesophase with negative anisotropy ($\chi_\alpha < 0$) will have the director perpendicular to H. In our system the smectic A phase of PA6ester1 block has a positive diamagnetic anisotropy, and so the mesogens' long axes are oriented parallel to the magnetic field H.

Fig. 4a shows the WAXS pattern obtained at T = 40°C on a CP4 sample aligned by a magnetic field. A typical aligned SmA structure is obtained. Small-angle Bragg reflections (I) are due to the smectic layers, and wide-angle reflections (II) are associated with molecular arrangements of mesogenic side groups within the smectic layers. Reflection at wide-angles (II) is preferentially located near the equator and indicates the liquidlike arrangement of the mesogens aligned macroscopically in a direction parallel to the magnetic field. The order parameter is estimated to be S = 0.74, through the angular intensity profile of wide angle reflection according to reported method.[30] The average distance between side groups is estimated to be 0.43 nm. Three orders of Bragg reflections (I) are visible along the meridian (parallel to the magnetic field). The smectic layer spacing associated with the Bragg reflections is p = 2.37 nm. This value is very close to that measured for the homopolymer (p = 2.52 nm).

In conclusion, even though the LC monomer is a monotropic nematic molecule, the LC homopolymer PA6ester1 possesses both nematic and smectic phases with the phase sequence: g-20°C-SmA-79.9°C-N-105.4°C- I. The block copolymer $PEG_{45}$-*b*-$PA6ester1_{20}$ (CP4) preserves the nematic and smectic phases in its LC block, with the phase sequence: g-15°C-SmA-60.8°C-N-73.9°C-I. The SmA phases in the homopolymer and in the block copolymer have very similar periods : p(PA6ester1) = 2.52 nm and p(CP4) = 2.37 nm. One-layer of anti-parallel packing is found for the SmA structure. Besides this basic molecular organization related to the mesomorphic properties of the LC block, what other type of self-assembled nanostructures, reflecting the incompatibility between the PEG block and the LC block, are formeed in this copolymer? In the following section we will examine this issue using SAXS at larger length scales.

**3.3. Self-assembly of $PEG_{45}$-*b*-$PA6ester1_{20}$ (CP4) in the pure state**

The sample aligned in a magnetic field was then studied by SAXS in another apparatus at room temperature in the absence of any magnetic field. The diffraction pattern is shown in Fig. 4b. Bragg reflections (I) correspond to the first order of reflections (I) (Fig. 4a) in a WAXS experiment and are due to smectic layers. The smaller angle reflections (III) are associated with the nanostructures of the microphase segregation of block copolymers. The smectic layer spacing by reflections (I) is p = 2.4 ± 0.1 nm, in good agreement with that found in WAXS. The signals (I) here are less oriented along the meridian than those for T = 40°C in a magnetic field. The absence of the magnetic field causes partial loss of the smectic alignment. Nevertheless, the block copolymer nanostructure is well aligned at room temperature as shown by the signals (III). Up to three orders of reflections are visible along the equator in Fig. 4b. These orders are associated with distances $d_1$ = 9.73 nm and $d_1/d_2/d_4 = 1/3^{1/2}/7^{1/2}$, which correspond to the 1$^{st}$, 2$^{nd}$ and 4$^{th}$ orders of a hexagonal phase. The hexagonal lattice parameter is $a_{hex} = 2d_1/3^{1/2}$ = 11.24 nm. When the copolymer composition is considered, the hexagonal structure must comprise cylinders of PEG and matrix of smectic polymer. The orthogonal orientation of reflection signals (I) from the smectic layers with respect to those (III) from the hexagonally packed cylinder planes indicates that the mesogens are parallel to the intermaterial dividing surface (IMDS) between the cylindrical nanodomains and the matrix. We infer then a homogeneous (planar) anchoring of mesogens at the IMDS. In Fig. 5 we show a schematic representation of the aligned self-assembled nanostructure of CP4 in a magnetic field.

The anchoring state (homogeneous or homeotropic) of the mesogens at IMDS and the sign of the diamagnetic anisotropy of the mesophase $\chi_a$, discussed in the preceding section, are two crucial parameters controlling the orientation of the copolymer nanostructures. These two parameters themselves are dictated by the chemical structure of the mesogens and the architecture of the block copolymers. The end-on side-chain block copolymer CP4 studied here combines positive diamagnetic anisotropy with homogeneous anchoring. The PEG cylinders consequently orient parallel to the field lines (1.7 T), yielding a monodomain with uniaxial symmetry, as shown in Fig.5.

Hamley et al.[14] reported an end-on side-chain LC block copolymer (Scheme 3) where the smectic mesophase also showed a positive $\chi_\alpha$, but with the difference that the mesogens had homeotropic anchoring at the IMDS of polystyrene (PS) cylinders. The application of magnetic field (1.8 T) failed to align the morphology of PS cylinders. This was ascribed to the nucleation of defects around the PS nano-cylinders in the LC matrix. In contrast Thomas et al.[16] described another end-on side-chain block copolymer (Scheme 4) with a negative $\chi_\alpha$ for mesophase and homogeneous anchoring of mesogens at the IMDS of polystyrene (PS) cylinders were found. A rather high magnetic field (9 T) still failed to produce long range order in the hexagonal cylindrical phase. As a matter of fact, the combination of homogeneous anchoring (mesogens parallel to the long axes of the cylinders) and the negative $\chi_\alpha$ (mesogens aligned perpendicular to H) results in a degeneracy in the orientation of the cylinders with respect to the magnetic field lines. Clearly, there are an infinite number of in-plane orientations in which the long axes of the cylinders can lie perpendicular to the magnetic field.

In conclusion, our block copolymer with the combination of positive diamagnetic anisotropy and homogeneous anchoring at the IMDS offers a good system for the formation of a long range ordered hexagonal cylindrical nanostructure. In the present study a capillary of 1mm diameter was used as container for block copolymer sample. The magnetic field for alignment was perpendicular to the capillary's long axis. The circular surface geometry of the capillary is not very favourable for the perpendicular alignment of mesogens along the magnetic field. Nevertheless, the orientation of the cylindrical phase and the smectic structure in the LC matrix was still clearly observed even at room temperature. We believe the alignment will be improved further for plane geometry where the block copolymer is in the form of a thin film on a substrate. As shown in Fig. 5, PEG cylinders are perpendicular to the surface along the magnetic field. If we change the magnetic field such that it is parallel to the surface, we would expect the cylinders to reorient parallel to the surface. This would constitue a magnetic field-triggered alignment of a self-assembled nanostructure.

**3.4. Self-assembly of PEG$_{45}$-*b*-PA6ester1$_{20}$ (CP4) in dilute aqueous solution**

Not only are the two blocks, PEG and PA6ester1, of the block copolymer incompatible in the pure state, but they also have very different affinity for water. The PA6ester1 block is hydrophobic, while the PEG block is hydrophilic. Self-assembly studies of CP4 in dilute water solution at room temperature were performed with the aid of the co-solvent dioxane. Fig. 6 shows the turbidity diagram when water is added progressively to dioxane solutions of the copolymer. The jump in turbidity upon the addition of water corresponds to the formation of particles which scatter light. The turbid mixtures at the end of the measurement (at around 40 wt% of water added) were dialysed against water to remove the dioxane and the particles suspended in pure water were then analysed by cryo-TEM.

Fig. 7 and Fig. 8 show typical cryo-TEM images of nanoparticles of CP4 embedded in ice. They are faceted vesicles with polygonal shapes, some of them exhibiting complex sharp-edge contours in several directions of the space (Fig. 7a). The vesicles are clearly unilamellar, with diameters ranging from 100 to 800 nm. The measured thickness of the hydrophobic part of the membrane is of order 10 nm. More interestingly, a careful analysis revealed a striped structure in some parts of the membrane (see Fig. 8 and the Fourier transform in the inset). The period measured is p = 2.5 ± 0.1 nm, in good agreement with that of the SmA phase measured in the pure sample of homopolymer and block copolymer CP4. We can conclude that the SmA structure of the PA6ester1 block is preserved in some parts of the vesicle membrane. Fig.9 shows a schematic representation of the smectic structure within a cross-section of the membrane where the stripes are visible.

Faceted vesicles have already been reported for small-molecular weight amphiphiles,[31,32,33,34,35,36,37,38] but not for polymer amphiphiles. It is known that planar bilayers in aqueous surroundings, or vesicle bilayers formed by small-molecular weight amphiphiles can exhibit several thermotropic substates. Above a certain melting transition, $T_m$, the amphiphile has fluid disordered alkyl chains and the bilayers appear in the fluid $L_\alpha$ phase. Below this temperature, upon the crystallization of the chains, gel-like phases (such as $P_{\beta'}$, $L_{\beta'}$ or $L_\beta$) are observed in the bilayers. Consequently, vesicles typically exhibit spheroidal shapes above $T_m$, while below $T_m$ nonspheroidal aggregates (e.g., disks, planar fragments, lens-shaped vesicles, regular polygon-shaped vesicles, and irregularly faceted vesicles),[31-35] can be formed because of the curvature constraints imposed by chain crystallization. In the case of faceted vesicles due to chain crystallization, reheating restores the spheroidal shape. In two-component catanionic vesicles or polymer associated catanionic vesicles, segregation has also been shown to occur in the frozen state and be responsible for the observed polygonal-shaped or faceted vesicles.[36,37] Only spheroidal shapes, however, have been observed up to now for polymer vesicles. Most of the polymer amphiphiles reported are coil-coil block copolymers, such as poly(ethylene oxide)-*b*-polybutadiene (PEO-*b*-PBD) or polyarylic acid-*b*-polystyrene (PAA-*b*-PS). Rod-coil block copolymers with nematic LC hydrophobic block were used to form polymer vesicles in our recent research.[21] We also observed spheroidal shaped vesicles for them. In the present work, the smectic LC structure of the hydrophobic block could be responsible for the faceted polymer vesicle shape. Work is in progress to study the vesicle shape evolution as a function of temperature (above the SmA-N transition and the N-I transition). Detailed investigations of the membrane structure and the topological defects of vesicles will also be carried out and described in a further paper.

## 4. Conclusions

In this paper we describe the synthesis and characterization of a series of amphiphilic LC block copolymers, in which the hydrophobic block is a smectic polymer poly(4-methoxyphenyl 4-(6-acryloyloxy-hexyloxy)-benzoate) (PA6ester1) and the hydrophilic block is polyethyleneglycol (PEG). The focus is on the self-assembly of one of the copolymers, $PEG_{45}$-b-$PA6ester1_{20}$ (CP4), in the pure state and in dilute aqueous solution.

In water, faceted unilamellar vesicles were formed by the copolymer CP4 as revealed by cryo-TEM. This is the first example of faceted shapes observed in polymer vesicles. In the lyotropic bilayer membrane, the thermotropic smectic order in the hydrophobic block is clearly visible in some places. The SmA layer normal is parallel to the membrane surface. Further investigations are necessary to elucidate the relationship between the smectic order in the membrane and the faceted shape of polymer vesicles.

Nano-structures in the pure state were studied by SAXS and WAXS. A hexagonal cylindrical micro-segregation phase was observed with a lattice distance of 11.2 nm. PEG blocks are located in cylinders, while the smectic polymer blocks constitute the matrix with a SmA layer spacing of 2.4 nm. The SmA phase has a positive diamagnetic anisotropy and the mesogens exhibit a homogeneous (planar) anchoring at the IMDS of PEG cylinders. The application of magnetic field aligns the SmA layer normal and the PEG cylinders parallel to the field lines, yielding a monodomain with uniaxial symmetry. This is a good system for forming long-range-ordered hexagonal cylindrical nanostructures. By changing the magnetic field direction, a magnetic field-triggered change in orientation of the nanostructure may be possible. We believe this system reveals important clues for the design of block coolymers with applications in nanolithography, high-density information storage media and organic photovoltaics.

**Acknowledgements**

MHL thanks Jacques Prost, Mark Bowick and Qiang Fang for fruitful discussions. We thank Aurélie Di Cicco for performing the cryo-TEM images and Lin Jia for his help in drawing the figures. We acknowledge financial support from the CNRS, the Institut Curie and the Agence Nationale de la Recherche (ANR-08-BLAN-0209-01).

**References**

† Bing Xu's present address is Shanghai Record Pharmaceuticals Co. Ltd, 799 Dun-Huang Road, Shanghai 200331, China (xubing97@hotmail.com).

‡ Rafael Piñol's present address is Fisica de la Materia Condensada, Facultad de Ciencias-ICMA, Universidad de Zaragoza-CSIC, Plaza San Francisco s/n, 50009-Zaragoza, Spain.

# Merveille Nono-Djamen's present address is Laboratoire Polymères, Colloïdes, Interfaces, UMR CNRS 6120, Université du Maine, 1 Av Olivier Messiaën, 72085 Le Mans Cedex 9, France.

Table and figure captions:

Table 1. Molecular characterization of the PA6ester1 and PEG-*b*-PA6ester1 series (CP1-CP5).

Table 2. Transition temperatures taken as the peak temperatures in DSC thermograms (°C) and enthalpies (in brackets) (J.g$^{-1}$) obtained by DSC analysis on heating at 10°C.min$^{-1}$.

Figure 1. Textures of the homopolymer PA6ester1 observed under polarizing optical microscope. (a) nematic textures at T=84.8°C. (b) SmA textures at T = 75.0°C.

Figure 2. DSC thermograms taken at 10°C.min$^{-1}$ : (A) and (B) the homopolymer PA6ester1 on heating and on cooling, (C) and (D) the block copolymer CP4 on heating and on cooling.

Figure 3. SAXS pattern of the homopolymer PA6ester1 at room temperature in the glassy SmA phase. The fibre sample was drawn from melted polymer, the long axis of the fibre being along the vertical direction. The smectic period is measured as p = 2.52 nm.

Figure 4. X-ray diffraction patterns of the block copolymer CP4. (a) WAXS pattern taken at 40°C *in-situ* after cooling from isotropic phase (85°C) at 0.1°C.min$^{-1}$ in a magnetic field of 1.7 T. (b) SAXS pattern taken at room temperature for the same sample in the absence of magnetic field. **H** is the direction of the applied magnetic field.

Figure 5. Schematic representation of the self-assembled nanostrucure of CP4 aligned in a magnetic field. The hexagonal lattice parameter is $a_{hex}$ = 11.2 nm, and smectic period is $p$ = 2.4 nm.

Figure 6. Turbidity diagrams of 1.0 wt% CP4 in dioxane upon addition of water.

Figure 7. Cryo-transmission electron micrographs of polymer vesicles of CP4 in water. Scale bar =100 nm.

Figure 8. Cryo-transmission electron micrographs of polymer vesicles of CP4. Scale bar =100 nm. The inset at higher left is an enlargement of the upper left area of the vesicle. The inset at lower right is the Fourier transform of the image, diffraction spots corresponding to a period of 2.5 ± 0.1 nm.

Figure 9. Schematic representation of the smectic structure within a cross-section of the membrane of CP4 polymer vesicles. The mesogens are represented by small elongated ellipsoids.

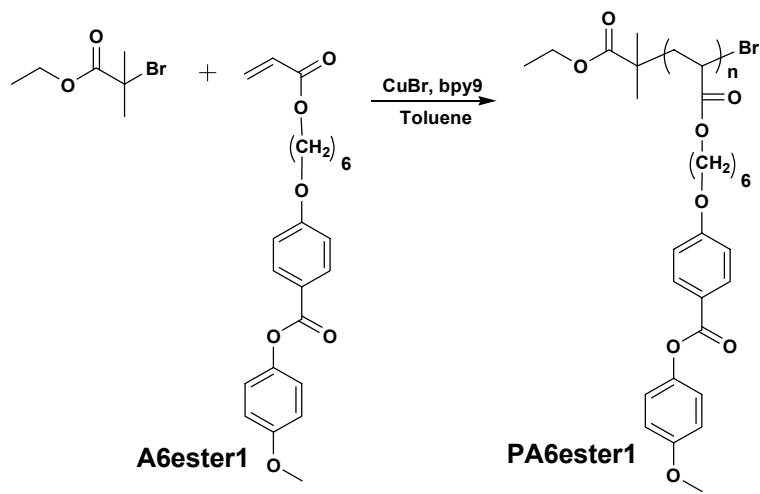

Scheme 1

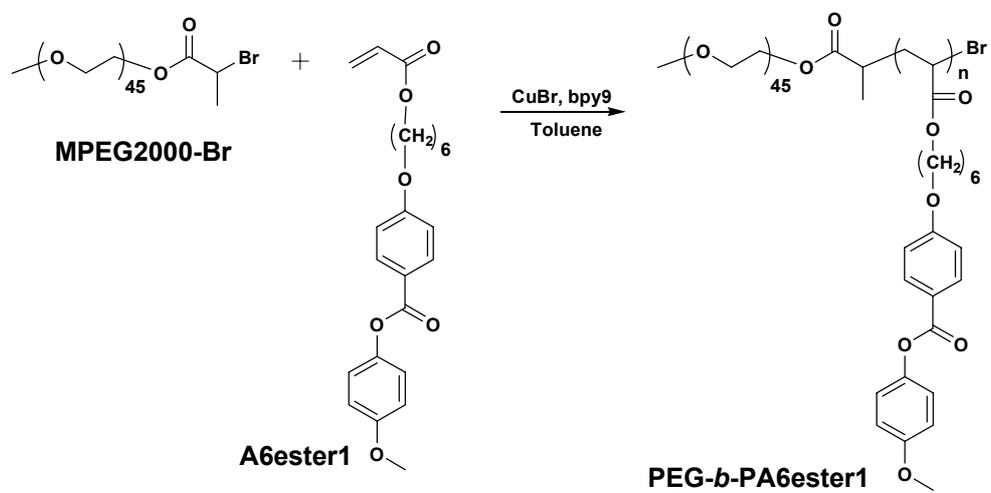

Scheme 2

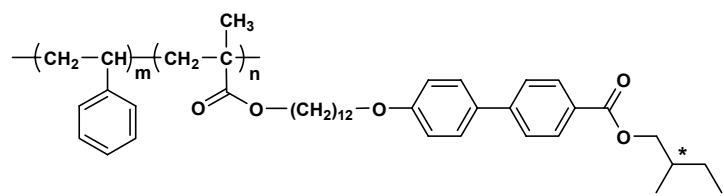

Scheme 3

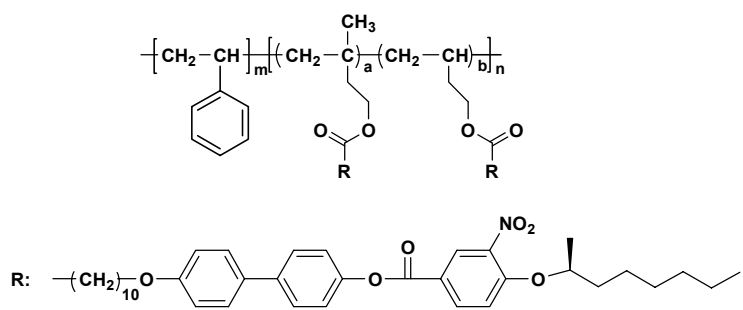

Scheme 4

Table 1

| Polymer | [M/[I]][a] | $M_n$ (Dalton) (NMR) | $M_w/M_n$ (SEC) | Hydrophilic/Hydrophobic weight ratio (NMR) | DP of PA6ester1 block $n$ (NMR) |
|---|---|---|---|---|---|
| PA6ester1 | 12/1 | 3300 | 1.17 | - | 8 |
| CP1 | 5/1 | 4000 | 1.07 | 50/50 | 5 |
| CP2 | 10/1 | 6000 | 1.09 | 33/67 | 10 |
| CP3 | 15/1 | 7900 | 1.13 | 25/75 | 15 |
| CP4 | 20/1 | 10000 | 1.20 | 20/80 | 20 |
| CP5 | 30/1 | 14200 | 1.32 | 14/86 | 31 |

[a] Molar concentration ratio of monomer to ATRP initiator. Reaction temperature: 80°C, reaction time: 24h.

Table 2

| Sample | Glass (g) | | Smectic A (SmA) | | Nematic (N) | | Isotropic (I) |
|---|---|---|---|---|---|---|---|
| PA6ester1 | ●[a] | 20 | ● | 79.9 (2.1) | ● | 105.4 (1.6) | ● |
| CP4 | ● | 15 | ● | 60.8 (1.5) | ● | 73.9 (0.3) | ● |

[a] The ● symbol means the phase exists.

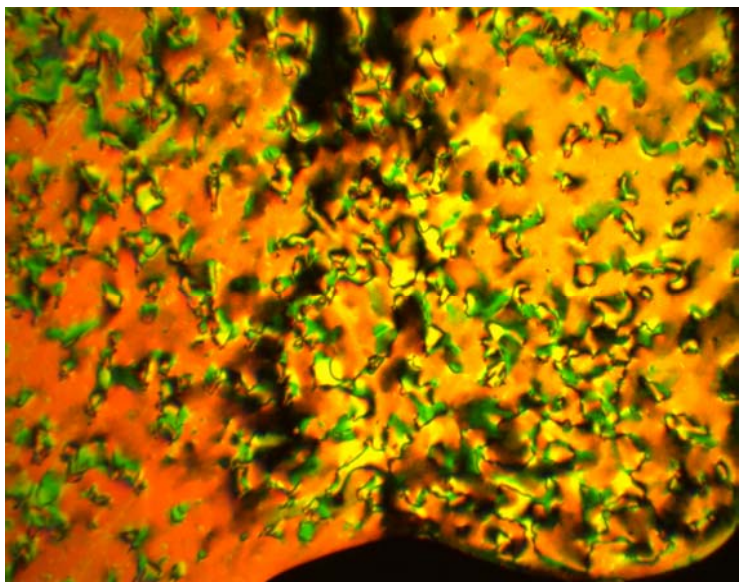

(a)

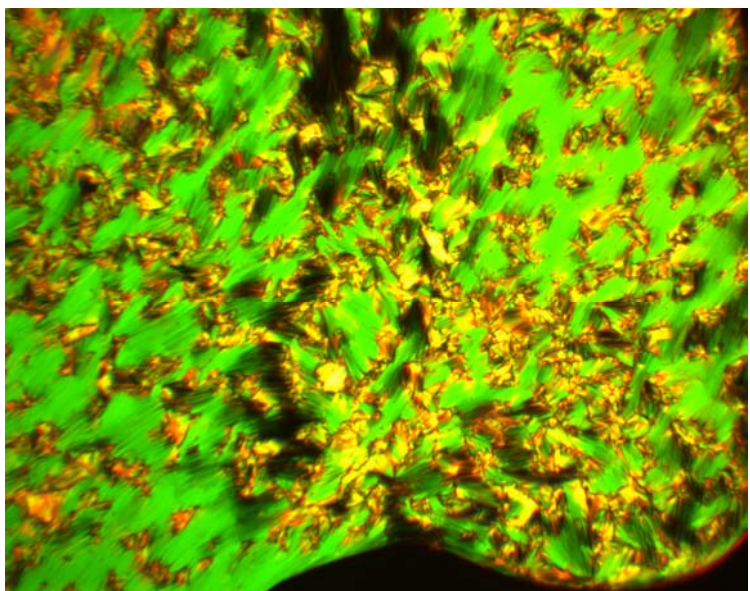

(b)

Figure 1

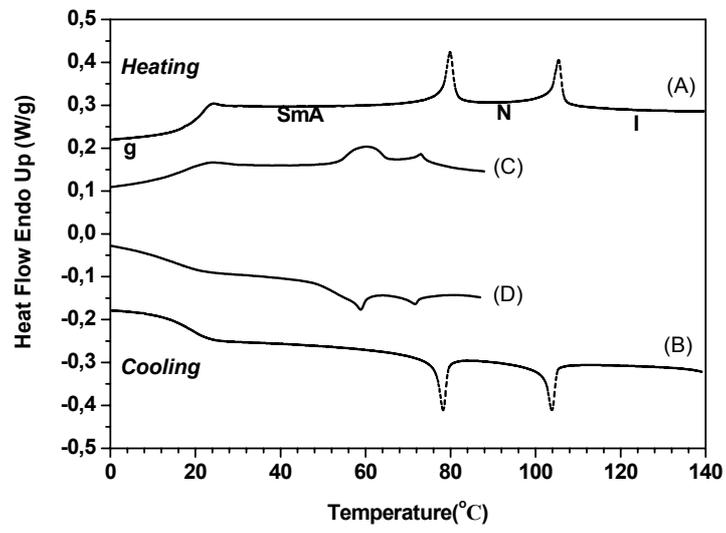

Figure 2

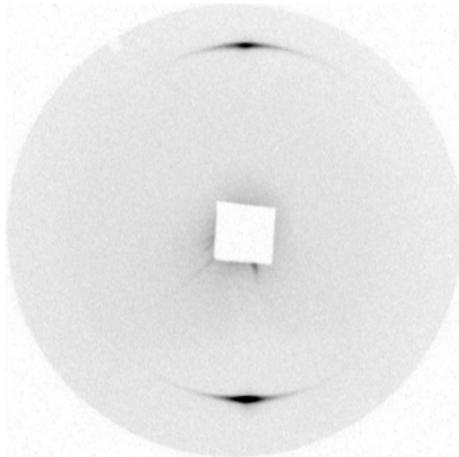

Figure 3

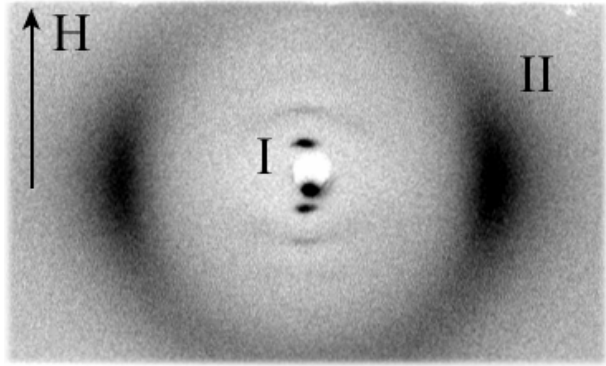

(a)

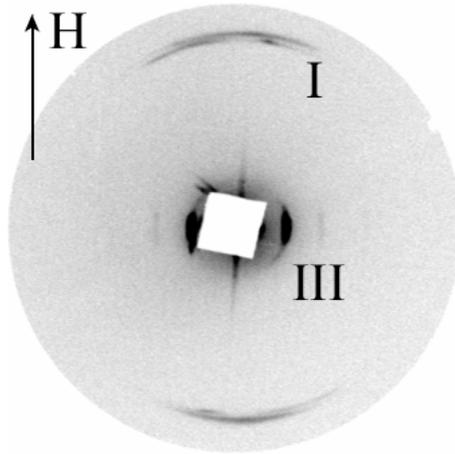

(b)

Figure 4

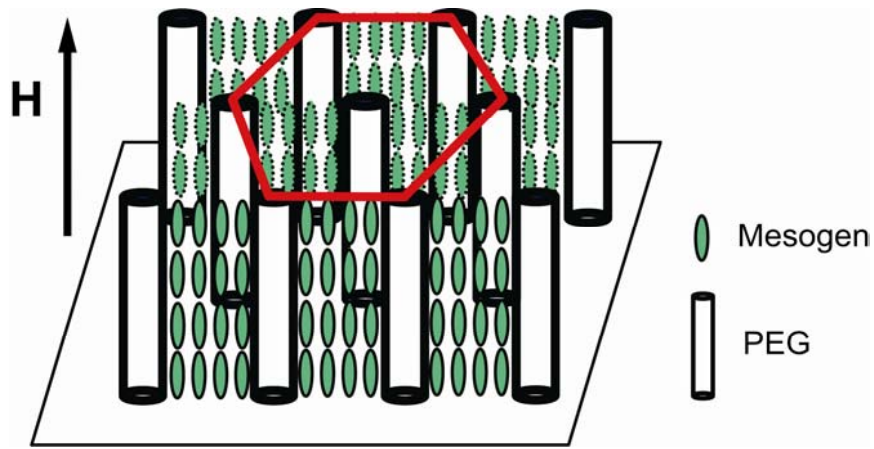

Figure 5

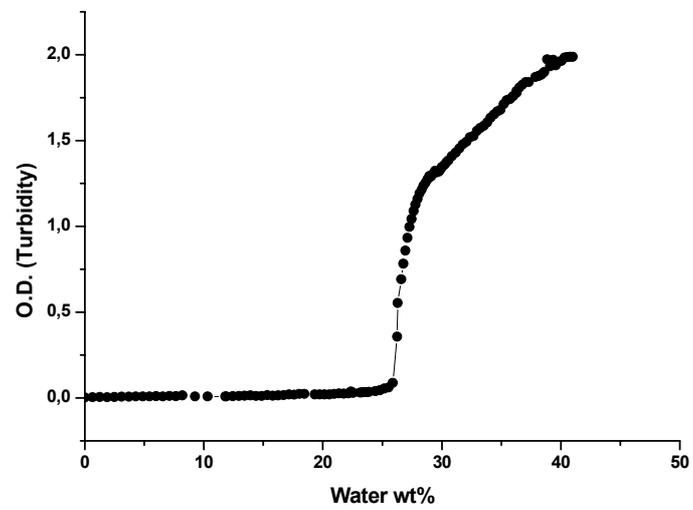

Figure 6

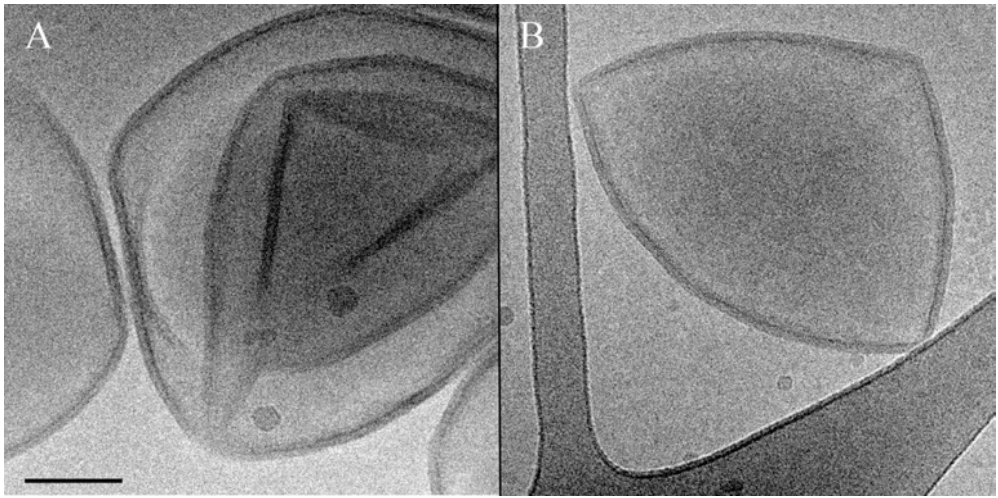

Figure 7

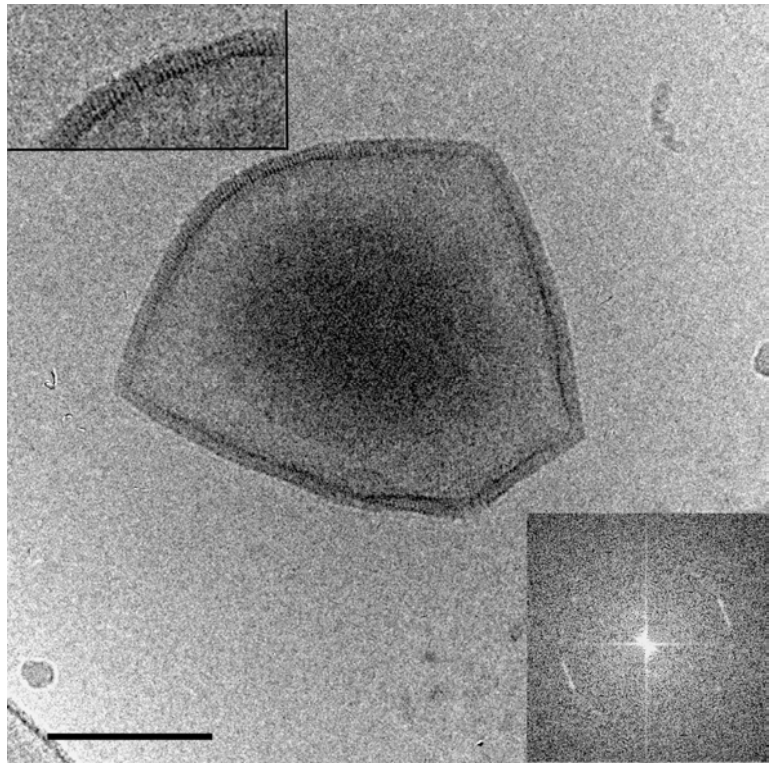

Figure 8

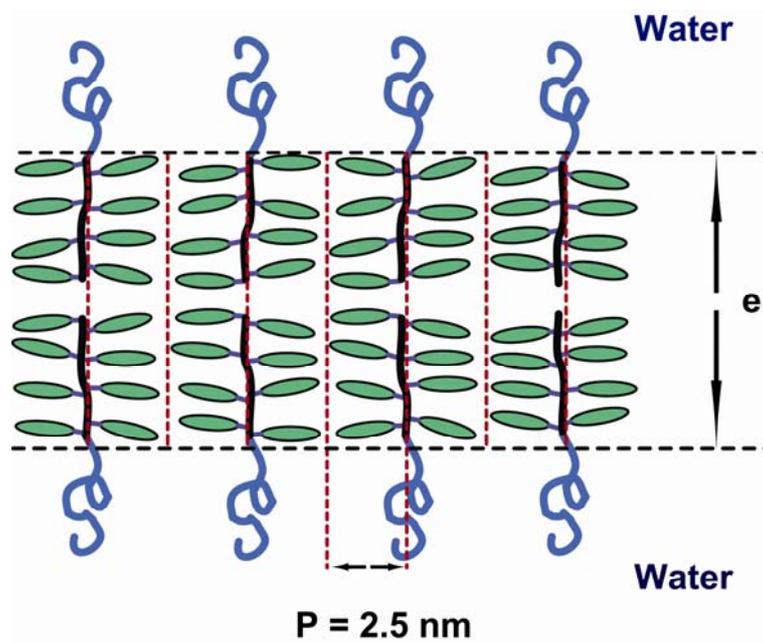

Figure 9